# Tunable dynamics of microtubule based active isotropic gels

Gil Henkin[1*], Stephen J. DeCamp[1*], Daniel TN Chen[1], Tim Sanchez[2], Zvonimir Dogic[1]

[1] Department of Physics, Brandeis University, Waltham, MA 02454 USA

[2] Department of Physics, Harvard University, Cambridge, MA 02438, USA

* these authors contributed equally to the work

Corresponding author contact: zdogic@brandeis.edu

Index Keywords: Active Matter, Gels, Bundled Microtubules, Molecular Motors, Spontaneous Flow

**Abstract:** We investigate the dynamics of an active gel of bundled microtubules that is driven by clusters of kinesin molecular motors. Upon the addition of ATP, the coordinated action of thousands of molecular motors drives the gel to a highly dynamical turbulent-like states that persists for hours and are only limited by the stability of constituent proteins and the availability of the chemical fuel. We characterize how enhanced transport and emergent macroscopic flows of active gels depend on relevant molecular parameters, including ATP, kinesin motor, and depletant concentrations, microtubule volume fraction, as well as the stoichiometry of the constituent motor clusters. Our results show that the dynamical and structural properties of microtubule based active gels are highly tunable. They also indicate existence of an optimal concentration of molecular motors that maximize far-from-equilibrium activity of active isotropic MT gels.

**Introduction**

The laws of equilibrium statistical mechanics impose severe constraints on the properties of conventional materials assembled from inanimate building blocks. Consequently, such materials cannot exhibit spontaneous motion or perform macroscopic work. Inspired by inherently far-from-equilibrium biological phenomena such as collective dynamics of a flock of birds, traveling metachronal waves of dense ciliary fields, or spontaneous streaming flows of a cellular cytoplasm, recent efforts have focused on recreating diverse biological functionalities in highly simplified materials [1-7]. Being recreated from the bottom up, the dynamical properties of these active materials are highly tunable, making them suitable for rigorously testing rapidly developing theoretical models of active matter [8-12], as well as their use in potential applications. In particular, recent experiments have shown that in the presence of attractive depleting interactions, short microtubule (MT) filaments assemble into 3D isotropic gels that exhibit persistent spontaneous flows, enhanced transport and self-mixing [13]. In contrast to conventional systems where flows are induced by inputting energy at the macroscopic scale, macroscopic flows of MT based active gels are driven by coordinated activity of thousands of microscopic kinesin molecular motors. We systematically vary relevant microscopic parameters of active gels and determine how these affect the emergent large-scale spontaneous flows. Our results demonstrate that the average speed at which the active gel spontaneous flows can be tuned by orders of magnitude. They also demonstrate that it is possible to systematically tune the spatial properties of active gels, such as the average vorticity lengthscale. Finally, they strongly suggest that there is an optimal concentration of molecular motors that maximize the emergent flows of active gels. Increasing the number of motors beyond this optimal number leads to internal frustrations that significantly slow down the overall gel dynamics.

**Active isotropic gels:** The three essential building blocks of active isotropic gels are filamentous MTs, clusters of kinesin molecular motors, and a depletion agent polyethylene glycol (PEG) (Fig. 1). MT filaments are stabilized by GMPCPP, a non-hydrolysable analog of GTP which suppresses dynamical instability and turnover of tubulin monomers [14]. GMPCPP also significantly lowers the nucleation barrier associated with MT formation, leading to an average filament length that is significantly reduced when compared to other methods of polymerization [15]. In our experiments the average length of GMPCPP stabilized MTs is 1.5 μm. Short MT filaments are essential for obtaining robustly flowing active gels. Longer MTs form highly elastic gels and the force provided by the motors is not sufficient to drive them to highly non-equilibrium states [16, 17]. We speculate that this is because longer stabilized MTs have greater propensity to crosslink with each other.

Adding non-adsorbing polymer such as PEG induces an attractive interaction between MT filaments by the well-studied depletion mechanism leading to the assembly of filamentous bundles [18, 19]. For all experiments described here, MTs are used at high concentrations at which they form a percolating 3D network which structurally resembles a classic viscoelastic gel-like material. However, unlike conventional gels active gels continuously undergo an ATP-driven dynamic network remodeling.

MT based gels are driven away equilibrium by clusters of kinesin molecular motors, which use chemical energy from ATP hydrolysis to take discrete 8 nm steps along the MT backbone [20, 21]. MTs are inherently polar filaments and the kinesin motors used in this study move specifically towards the MT plus end. Kinesin motors are labeled with a biotin linker and a tetrameric streptavidin binds several motors into multi-motor clusters (Fig. 1b). Such clusters can simultaneously bind and translocate along multiple MT filaments and can drive an isotropic MT suspension into a variety of defect ridden states that are inherently far-from-equilibrium [5-7]. We expect that the activity of the motor induced filament sliding is significantly enhanced for MTs in a bundled geometry when compared to filaments that are on average isotropic. On average a single kinesin motor moves about a micron along a filament before detaching [20]. The effective motor processivity is significantly enhanced for clusters moving along bundled MTs. Even after one motor within the cluster unbinds, the cluster can remain attached to the bundle through other motors in the cluster. The bundled geometry then presents multiple binding sites to which the detached motor can quickly reattach, thus enhancing motor processivity and overall activity.

The nature of the sliding force depends on the relative MT polarity in the bundle (Fig 2a) [5]. When adjacent MTs are polarity aligned, kinesin motors will jointly walk towards the plus ends of the bundle without generating interfilament sliding [5]. In contrast, when two MTs are anti-aligned (MT plus ends are on distal ends), kinesin motor clusters will move towards opposite ends and cause the filaments to slide past one another. When acting on an isolated bundle the kinesin clusters will locally polarity sort the MT filaments [22, 23]. Once the bundle is polarity sorted it remains in a quasi-static configuration as motors simply translocate along the polarity sorted regions without inducing any further filament sliding.

Experimentally, we observe that such polarity sorting is frequently accompanied by the overall bundle extension (Fig. 2b) [13].

The dynamics of dense percolating 3D MT bundled gels driven by kinesin clusters is considerably more complex than that of dilute isolated bundles. Similar to isolated bundles, kinesin clusters induce local MT polarity sorting. However, since each polarity-sorting segment is connected to a background network, local extension leads to buckling of the composite bundled structure (Fig. 2c). Once the local curvature reaches a high enough magnitude the composite filamentous bundle fractures, producing shorter fragments. The typical buckling lengthscale is about a hundred microns, which is significantly larger than the length of the constituent one-micron-long MTs. This demonstrates that bundle disintegration does not fragment individual MTs but only serves to break up the composite bundles held together by weak depletion interactions. Fragments produced due to self-driven buckling are quickly incorporated into the background network with random polarity. At this point the molecular motor once again drives local polarity sorting and extension of constituent bundles. Therefore, the microscopic dynamics of active isotropic gels are driven by repetitious cascades of MT bundle polarity sorting, extension, buckling, fragmentation, and subsequent recombination with the background network. The molecular motors continuously sort the MTs but due to the buckling instability the entire gel never attains a global polarity-sorted quiescent state.

The microscopic speed at which kinesin motors move along a MT backbone can be tuned by varying ATP concentrations [21]. However, simply dissolving a fixed initial amount of ATP in an active gel would lead to an undesirable situation where the motors continuously deplete the chemical fuel leading to gel dynamics that monotonically slows down for the entire duration of the experiment. Such conditions would make it impossible to prepare samples that have persistent steady state microscopic dynamics over multiple hours required for quantitative studies. To overcome this challenge, we incorporate an ATP regenerating system into our active gel [24]. As soon as a kinesin motor hydrolyzes an ATP into an ADP, pyruvate kinase/lactic dehydrogenase enzymes (PK/LDH) use energy from a high-energy chemical compound phosphoenol pyruvate (PEP) to regenerate ATP. Using such a regeneration system, it is possible to keep a fixed ATP concentration that ranges from micromolar to millimolar for sufficiently long time to perform quantitative measurements. In practice we find that MT gels with ATP regeneration can remain active for multiple days. The limiting factor appears to be protein stability as the MTs and kinesin motors start to denature at room temperature after approximately 2 days.

**Materials and Methods**

**Polymerization of MTs:** Tubulin was purified from bovine brains with two cycles of polymerization/depolymerization in 1,4-piperazindiethanesulphonic (PIPES) buffer [25]. MTs were polymerized at concentrations of 6-8 mg/ml in an M2B buffer (80mM PIPES, 1mM EGTA, 2mM MgCl2) using Guanosine-5'[($\alpha,\beta$)-methyleno]triphosphate (GMPCPP), a non-hydrolyzable analog of GTP. GMPCPP lowers the nucleation barrier and leads to polymerization of short MTs that have an average length of 1.5 $\mu$m [15]. In some experiments, MTs were labeled using Alexa 647 dye which was attached by a succinimidyl ester linker to a primary amine on the tubulin surface [26]. Tubulin suspension polymerized by incubating at 37 $^o$C for 30 minutes. Subsequently, it was allowed to sit at room temperature for an hour. Subsequently, the MT suspension was frozen at -80 $^o$C. Repeated freezing and thawing did not alter the MT length distribution (average 1.5 $\mu$m MTs). For experiments that required varying filament concentration, polymerized MTs were centrifuged for 10 min at 80,000 RPM (Beckman Optima TL, TLA100.4). Supernatant was aspirated and the MTs were resuspended in M2B buffer at concentration of ~90 mg/ml. The highly concentrated MT suspension was diluted to a final desired concentration and used within a day of polymerization.

**Kinesin-streptavidin motor clusters**: K401-BIO-6xHIS is the 401-amino acid N-terminal domain derived from *Drosophila melanogaster* kinesin and fused to the *Escherichia coli* biotin carboxyl carrier protein expressed and purified from *E. Coli*. The kinesin is also labeled with a six-histidine tag [27]. Kinesin motors were purified on a nickel column dialyzed against 50 mM imidazole, 4mM MgCl2, 2mM DTT, 50μM ATP, and 36% sucrose buffer, flash frozen in liquid nitrogen and kept at -80 °C. Thawed kinesin was mixed with tetrameric streptavidin at varying ratios in an M2B buffer solution that was supplemented with DTT. Kinesin clusters can be frozen and thawed multiple times without significant change in system activity.

**PLL-PEG coated tracer beads**: To suppress depletion induced binding to network backbone, micron sized polystyrene beads where coated with a polymer brush. Briefly, poly-L-lysine - polyethylene-glycol block copolymer was synthesized by conjugating a N-hydroxysuccinimidyl ester of methoxypoly(ethylene glycol) propionic acid with poly-G-lysine in 50 mM sodium borate buffer, pH 8.5. The poly-lysine backbone has a positive charge that binds strongly to negatively charged polystyrene beads, leaving a brush of neutral PEG covering the bead surface. Labeling the beads was done by mixing 10 μL PLL-PEG, 100 μL HEPES (7.5 pH) buffer, 12 μL stock 3 μm polystyrene beads, and 878 μL H2O. Solution was incubated for one hour on a rotating platform to prevent bead settling, sonicating once every 20 minutes. After incubating, the beads were spun for 5 minutes at 4000 rpm, consolidated in a tube with M2B buffer, spun again, and resuspended in 50 μL M2B.

**Sample chambers**: To suppress depletion of MTs onto surfaces, sample chambers were coated with a polyacrylyamide brush [28]. Coverslips and slide glass were sonicated in 0.5% detergent solution, followed by an ethanol bath, and finally a 0.1 M KOH bath to clean and etch the glass. Slides and coverslips were then immersed in a mixture of 98.5% ethanol, 1% acetic acid, and 0.5% 3-(trimethoxysilyl) propylmethacrylate and allowed to soak for 15 minutes so silane could coat the glass surface. Slides and coverslips were rinsed with DI water before being immersed in 2% acrylamide monomer solution, 35 μL per 100 mL solution of tetramethylethyldiamene (TEMED) and 70 mg per 100 mL of ammonium persulfate were added to induce formation of polymer brush. Slides and cover slips were kept in solution for up to two weeks. Before use they were thoroughly rinsed with DI water and dried.

**Protocol for assembly of active gels**: Active pre-solutions (without MTs) for active gels contained ATP, anti-oxidizing agents and Trolox to avoid photo-bleaching, an ATP regeneration system based on phosphoenol pyruvate (PEP) and pyruvate kinase/lactic dehydrogenase (PK/LDH), and depletant polymer poly-ethylene glycol (PEG, 20 kDa). All components were suspended in M2B buffer. The active pre-solution was prepared by dissolving: 1.33 μL anti-oxidant stock (15 mg/ml glucose and 2.5M DTT), 1.33 μL anti-oxidant stock (10 mg/ml glucose oxidase and 1.75 mg/ml catalase), 1.7 μL PK/LDH, 2.9 μL high-salt (68 mM MgCl2 in M2B) buffer, 6 μL trolox (20mM), and 8 μL PEP (200mM). Appropriate amounts of M2B, kinesin-streptavidin clusters, and PEG were added to the active pre-solution in accordance with the experiment being performed. The total volume of the active pre-solution after mixing components was 40 μL. Final active gels were made with 4 μL of active pre-solution, 1 μL tracer bead solution, and 1 μL MTs. The standard active gel included 3.9 μL of the kinesin-streptavidin cluster mix, 8 μL of PEG, and 5.14 μL M2B in the pre-solution before mixing with MTs and beads.

To ensure sample-to-sample uniformity and reproducibility, we prepared a large pre-solution mixture for each experiment. Samples were observed in flow cell chambers that were 18 mm long, 2 mm wide, and 100 μm deep. Data was taken at the midplane in order to minimize hydrodynamic interactions with boundary surfaces. Using fluorescence microscopy, the active gel was monitored until it reached stable activity (Fig. 3). Beads embedded in the active gel were imaged using brightfield microscopy with a 10x objective at frames rates that varied from 1 frame/sec to 10 frames/sec, depending on the apparent network speed. Bead tracking was performed using custom MATLAB code based on algorithms described by Crocker and Grier [29]. From the tracked bead positions, complete particle trajectories were constructed.

Each set of experiments varied one parameter while keeping all other parameters constant. Unless otherwise indicated ATP was fixed at 1.43 μM, MTs at 1 mg/ml, active cluster concentration at 0.41 μM (mixed at a ratio of 1.7 kinesins per streptavidin) and depleting agent at 0.8% PEG (w/w).

**Results**

To quantify flow dynamics we suspended micron-sized beads into active gels. Being coated with a polymer brush the beads do not adsorb onto the MT backbone, but are advected by the spontaneous flow generated by the active isotropic gel (see ESM Movie 1). Quantifying bead trajectories provides information about the spatiotemporal characteristics of active flows. First, we have determined how the dynamics of emergent flows depends on ATP concentration which controls the speed at which kinesin motor clusters move along the MT [21]. The measured mean-squared displacement (MSD) of the tracer beads was highly dependent on ATP concentration (Fig. 4a). For very low ATP concentration, the beads were trapped by an enveloping MT network and they exhibit sub-diffusive MSD curves, a behavior that is characteristic of conventional viscoelastic gels [30]. With increasing ATP concentration the slope of the MSD curves gradually increased. At these intermediate ATP concentrations the MSD curves transition from sub-diffusive to super diffusive. In the transition region the slope of the MSD curves is larger on longer time scales. For ATP concentrations above 28 μM, MSD curves become ballistic, indicating that the beads move essentially along straight trajectories on experimental timescales. At even longer time scales the MSD curves will become diffusive. However, we do not have the capability to track beads for sufficiently long time to explore this regime.

Next we examined how the average network speed, defined as the average distance the beads move over 10 seconds, depends on the ATP concentration (Fig. 4b). Single molecule experiments revealed that kinesin motor speed depends on ATP concentration according to Michaelis-Menten kinetic model. In this model the kinesin velocity is given by $v = V_{max}[ATP]/([ATP] + K_m)$ where $V_{max}$ is kinesin velocity at saturating ATP concentration and $K_m$ is the Michaelis constant [31]. We observe that the network speed monotonically increases with increasing ATP concentration before saturating. This is in qualitative agreement with measurements for single motors. However, we note that the characteristic network speed cannot be quantitatively fitted to the Michaelis-Menten kinetics. Furthermore, for saturating conditions the average network speed is 2 μm/sec, which is significantly faster than $V_{max}$ of single kinesin which saturates at 0.8 µm/sec. A plausible reason is that motion of multiple extensile segments within a single bundle add up to velocities that are significantly larger than the motions of individual molecular motors.

MSD curves provide information about temporal bead dynamics. To characterize spatial variations of spontaneous flows we have also measured normalized equal-time spatial velocity correlation functions (VCF), $\frac{<v(r+\Delta r)v(r)>}{<v(r)>^2}$, where $v(r)$ is the velocity of a background fluid at position r (Fig. 4c). The measured correlation functions decay monotonically with increasing bead separation $\Delta r$. After decaying to zero they frequently become negative, indicating local rotation or vorticity within the fluid flow. Sometimes, multiple minima at larger separations are present due to multiple vortices in the field of view. From the decay of the velocity correlation function it is possible to define the characteristic spatial flow lengthscale, λ associated with the emergent flow. With increasing ATP concentration, λ increases slightly before saturating (Fig. 4d). Our current hypothesis is that λ is closely linked with the buckling lengthscale of the constituent MT bundles. However, to confirm this it is necessary to directly visualize dynamics of the constituent 3D bundle network, something that has not been accomplished so far.

Besides MSD and VCF we have further analyzed the dynamics of active gels by measuring the probability density function (PDF) of the tracer particle displacements at time t given that at time t=0 the particle is located at origin. The PDF of thermally diffusing beads with no confinement is Gaussian. For conventional gels whose mesh size is comparable to the tracer size, one expects Gaussian behavior for displacements smaller than the mesh size, but occasional ''jumps'' larger than the mesh size, leading to PDF that has broader tails than those predicted by a Gaussian distribution [32]. For MT gels at zero ATP concentration we measure a similar PDF, demonstrating that they behave as conventional gels (Fig. 5b insert). We quantitatively characterize the degree of Gaussianity by the distribution's excess kurtosis: $<u^4>/<u^2>^2$ - 3, where $<u^4>$ is the fourth moment and $<u^2>$ is the variance. Defined in this way, a Gaussian distribution will have an excess kurtosis of zero. An excess kurtosis of less than zero represents a distribution that has a flatter peak and narrower tails than a Gaussian. An excess kurtosis of greater than zero represents a distribution that has a sharper peak and wider tails than a Gaussian. With increasing ATP concentration we observe a systematic trend where the Gaussian tails disappear and become narrower than Gaussian (Fig. 5a). Consequently the excess kurtosis, which is initially larger than zero, quickly crosses zero and becomes negative (Fig. 5b).

We have also examined how the PDFs change with increasing time interval, t at fixed ATP concentration (Fig. 5c). We see two limiting behaviors depending on the activity regime. In the equilibrium regime with zero ATP, the system exhibits 'mesh jumps' which have broader than Gaussian tails for all times [32] (Fig. 5d). For non-equilibrium regimes at higher than 56 µM ATP concentration the mesh is no longer well-defined, leading to an effectively Gaussian PDF for short times. This converges to a narrower than Gaussian PDF at long times, consistent with the concentration behavior. We find the same generic behavior for all gel parameters that we vary, i.e. a distribution that is narrower than Gaussian (negative excess kurtosis) at long times.

In suspensions of biological swimmers, one typically observes a displacement PDF that is well described by a Gaussian core with additional broader tails [33-36]. The Gaussian core arises from the thermal motion, while the hydrodynamic coupling to background swimmers leads to broader-than-Gaussian tails. The broad tail is quantified by a positive excess kurtosis which converges towards Gaussian as concentration of swimmers increases due to the superposition of many incoherent flow fields, which mimics random Brownian motion. Our system is fundamentally different from suspensions of swimmers in that we have a negative kurtosis. This negative kurtosis is manifested as a narrower-than-Gaussian tail. We hypothesize that the narrow distribution is a consequence of having relatively few active components which are large extended objects, in contrast to having many small, independent active components as in the case of swimmer suspensions. The deformation of active bundles drive large-scale flows and thus coherently moves many tracer particles with roughly the same speed, resulting in a narrower-than-Gaussian PDF.

Next we have determined how active gel dynamics depends on the concentration of kinesin motor clusters. While the MSD curves exhibited ballistic behavior irrespective of the motor cluster concentration (Fig. 6a), the average gel speed exhibited interesting dependence on the same parameter. Increasing the motor cluster concentration up to 0.5 µM increased the network speed (Fig. 6b). At this point there are approximately 11 motor clusters per each MT filament. However, increasing motor cluster concentration beyond this point slowed down the network dynamics. This observation demonstrates that the fastest gel dynamics occurs for an optimal number of motors per each MT. The characteristic flow lengthscale $\lambda$ also exhibited pronounced dependence on motor cluster concentration (Fig. 6c-d). At low motor concentrations, $\lambda$ was large and it steadily decreased with increasing motor concentrations.

It is easy to understand why the network speed increases with increasing cluster concentration in the low concentration limit. For very low cluster concentration most of the adjacent MTs will not be crosslinked by active motor clusters. Therefore, the overall extensile network dynamics, being driven by a low volume fraction of actively sliding MTs pairs will be quite slow. Increasing the cluster density increases the fraction of sliding MTs pairs and thus the overall network dynamics. However, the existence of an optimal concentration of motor clusters beyond which the average speed monotonically MTs slow down at high motor density for tightly coupled motors with short stalks, while for loosely coupled motors with longer stalks the MT velocity is essentially independent of the number of engaged motors [37, 38]. It is also possible that in the high motor density limit there are too many kinesins per MTs which might lead to ''traffic jams'' as has been seen in kinesin-8 motor systems [39, 40].

Next we have studied how the active gel dynamics depends on MT concentration (see ESM Movie 2). In these experiments we increase MT concentration from 1 mg/ml to 25 mg/ml, while keeping the concentration of kinesin clusters and ATP constant. We found that the MSD curves displayed ballistic behavior for all MT concentrations examined (Fig. 7a). However, the average speed of the spontaneous flows exhibited a complex non-monotonic dependence on MT concentration (Fig. 7b). Initially the average speed increased with increasing MT concentration reaching a maximum value of 4 µm/sec for MT concentrations of 5 mg/ml. Increasing filament concentrations beyond this value gradually slowed the gel dynamics. Intriguingly, changing the MT concentration by more than an order of magnitude did not significantly alter the characteristics spatial flow lengthscale $\lambda$ (Fig. 7c-d). The existence of an optimum MT concentration is in qualitative agreement with the experimental observations in which we varied kinesin cluster concentration (Fig. 6). However, for varying MT concentration the optimal ratio of motor clusters per MT is ~55 which is significantly different from the optimum concentration found in figure 6. At high MT concentration there are very few motor clusters per MT and thus the overall dynamics is very slow. With decreasing MT concentration the motor cluster/MT ratio increases leading to increased speed of the overall network dynamics. Once past the optimal ratio of clusters/MT, there are too many molecular motors per MT and the overall dynamics slows down, similar to what is observed in Fig. 6.

Another parameter that critically affects the emergent dynamics of active networks is the concentration of the depleting agent (see ESM Movie 3). Non-adsorbing polymer, PEG, induces attractive MT interactions and their lateral association into filamentous bundles (Fig. 1a). In the depletion picture the strength of the attractive interactions is simply controlled by the PEG concentration. Below PEG concentrations of 0.6%, depletion is not strong enough to induce bundle formation and the sample remains in an isotropic state. In contrast to previous work that demonstrated formation of MT asters in mixtures of kinesin clusters and isotropic MTs [6, 7], adding motor clusters to isotropic suspensions of GMPCPP MTs does not induce any large-scale dynamics. This is presumably because the average length of GMPCPP MTs is much shorter than those stabilized by taxol. Increasing the depletant concentration to 0.8% induces MT bundling and network formation which is accompanied by the emergence of spontaneous flows characteristic of active gels (Fig. 8a). The transition from non-bundled largely static gels at low depletant concentrations to highly dynamic gels at larger depletant concentration is very sharp. Intriguingly this is the point at which the gel exhibits the fastest dynamics. Increasing PEG concentration beyond this value slows down the average flow speed (Fig. 8b). The characteristic speed at 0.8% PEG was 2 μm/sec; by increasing PEG concentration to 4.4% the gel slows down to 0.3 μm/sec. This trend can be perhaps understood by considering that stronger MT attractions (highest depletant concentrations) lead to increased friction associated with interfilament sliding of adjacent MTs. In this case the motor clusters working against an enhanced load due to larger sliding friction would move at slower speeds. However, very little is known about frictional coupling between bundled filaments and how it depends on depletant concentration. The spatial flow lengthscale, λ, also exhibited strong dependence on PEG concentration, monotonically decreasing from 150 μm to 75 μm with increasing depletant concentration (Fig. 8c-d.).

Finally, we varied the ratio of kinesin motors that are bound to tetrameric Streptavidin (see ESM Movie 4) (Fig. 9). While increasing the ratio of kinesin to streptavidin, the characteristic quantities behave strikingly similar to the motor concentration dependence. This is not surprising as in both cases we are manipulating the ratio of the number of kinesin motors per MT. The peak in the characteristic speed of the system occurs with a molar ratio of 1.7 kinesin per Streptavidin (Fig. 9b), however, it should be noted that while the molar ratio suggests cluster sizes of about 2 motors it has been shown that clusters with upwards of 8 motors are possible due to the presence of 2 biotin labels on each kinesin dimer [6, 7].

**Discussion**

Active MT gels are intriguing far-from-equilibrium materials that exhibit macroscale spontaneous flows, enhanced transport, and mixing that is driven by the coordinated motion of thousands of constituent microscopic molecular motors. All the essential microscopic ingredients of MT based active gels such as the depletion interactions, the mechanics of the constituent MT filaments, or the stepping of kinesin motors have been characterized with great quantitative detail [19, 41, 42]. Therefore, it should be feasible to formulate theoretical models that establish a quantitative link between relevant microscopic parameters and the emergent macroscopic dynamics. However, we note that such theoretical models have not been developed so far and therefore it remains a significant challenge to explain our experimental data. At mesoscopic lengthscales these theoretical models should also account for buckling and fragmentation of MT bundles which are held together by soft depletion interactions. These events are essential for generation of spontaneous flows at macroscopic scales. However, the mechanics of composite filamentous bundles remains a relatively unexplored area, and only recently, theoretical efforts have focused on describing certain aspects of their properties [43, 44].

Being assembled from a few well-characterized components the properties of active gels are highly controllable. For example, by controlling the ATP concentration we tuned the speed of constituent kinesin motors from 0.1 μm/sec to 1 μm/sec. Furthermore, by using an ATP regeneration system, we held this motor speed constant over multiple hours allowing us to prepare and quantitatively characterize true steady state active materials. These unique capabilities of active isotropic gels have allowed us to tune the enhanced diffusion of tracer particles over many orders of magnitude. Such enhanced diffusion is reminiscent of measurements of tracer particles in active systems that are driven by living microswimmers such as *Escherichia Coli* or *Chlamydomonas reinhardtii* [34, 45, 46]. However, active systems based on living organisms are not easily tunable and consequently the MSDs of tracer particles suspended in a bath of living swimmers usually exhibit only one type of behavior. In contrast, the MSD curves of tracer beads suspended in active MT gels could be tuned from sub-diffusive to super-diffusive and ballistic behavior. The divergence of an active gel from a bath of living swimmers also manifests in the kurtosis of displacement PDFs, which tend to be negative in MT gels, indicating that the flows that arise have more coherency.

Unfortunately, the parameters that we are able to tune in the bundled active MT system are not orthogonal parameters, i.e. changing one experimental parameter simultaneously changes multiple properties of active gels such as elasticity of the constituent bundle as well as their activity. For example, tuning ATP controls the speed at which Kinesin motors walk along MTs and one would naively expect it to only affect the activity. However, ATP concentration also affects the motor processivity and as such it influences network elasticity. Motors act as rigid crosslinkers at low ATP concentrations, resulting in stiffer bundles. This makes even quantitative interpretation of our results difficult. Perhaps the most robust effect we have measured is a well-defined number of motors that optimizes the overall dynamics of active gels. Increasing motor concentration beyond this limit slows down overall network dynamics (Fig. 6b, 7b, 9b).

Finally, we note that active isotropic gels are built from MT bundles, which are predominantly extensile in character. This is in sharp contrast to the other major class of biologically inspired active gels that are built from actin filaments and myosin motors, and which are always contractile [47-49]. The reason for this difference is not well understood, although it has been suggested that actin gels are contractile due to small wavelength associated with buckling instability of the constituent actin filaments. MTs being orders of magnitude stiffer that actin filaments have a much longer buckling wavelength and this might suppress generation of any contractile forces [3, 50].

**Acknowledgments**

This work was primarily supported by Department of Energy, Office of Basic Energy Sciences under Award DE-SC0010432TDD. Acquisition of preliminary data as well as the development of the analysis software was supported by the W. M. Keck Foundation and NSF-MRSEC-0820492. We also acknowledge use of Brandeis MRSEC optical microscopy facility (NSF-MRSEC-0820492).**Electronic Supplementary Materials (ESM)**

**Movie 1. Tracer particles suspended in active gel.** The activity of the gel generates spontaneous flows which advects 3 μm tracer beads suspended in the background fluid. The beads are imaged under brightfield microscopy and particle tracking is performed to construct bead trajectories.

**Movie 2. Active gel with varying MT concentration.** Under fluorescence microscopy, the active gel is imaged over a range of MT concentrations (1mg/ml – 25mg/ml). We find that there is an optimal number of motors per MT with maximizes gel dynamics.

**Movie 3. Active gel with varying PEG concentration.** The depletant agent, poly-ethylene glycol (PEG), varies the strength of the effective attractive interaction between MT filaments. Under fluorescence microscopy, the effect of varying PEG can be seen in the resulting morphology of the active gel for two concentrations of PEG (0.8% w/w and 4.4% w/w).

**Movie 4. Active gel with varying the ratio of kinesin to Streptavidin.** Changing the stoichiometric ratio of kinesin to Streptavidin effectively alters the resulting size the kinesin motor clusters. Under fluorescence microscopy, the effect of varying the ratio of kinesin to Streptavidin induces a morphological change in the active gel as seen by the thickness of MT bundles and a change in the characteristic speed of the gel. We find that there is an optimal number of motors per MT with leads to the fastest gel dynamics.

**Figures and Figure captions**

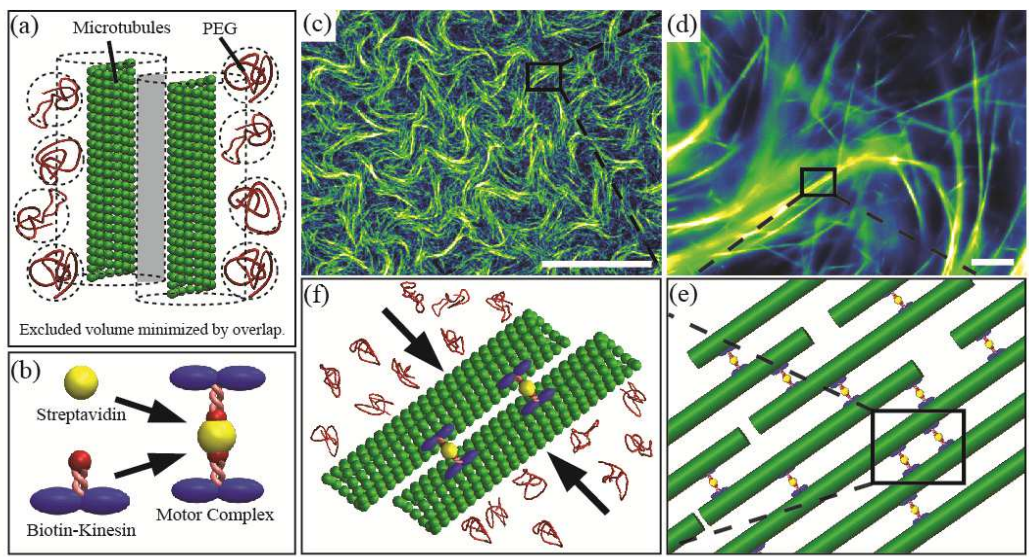

**Figure 1. Deconstruction of the bundled active MT gel. a)** The active gel is composed of MT filaments which are bundled together through the depletion interaction using PEG. The depletion interaction serves to minimize the volume in the system which excludes the center of mass of the PEG by aligning and bundling filament along their long axis. **b)** Kinesin motor complexes are assembled from biotinylated kinesin motor proteins linked together by tetrameric streptavidin. **c)** A large field of view of the active gel under fluorescence microscopy shows large scale bending of MT bundles. Scale bar is 250µm. **d)** Zooming into a region of the gel reveals a gel structure of bundles of polydisperse size. Scale bar is 10µm. **e)** A schematic of a MT bundle depicts many MT filaments crosslinked by the Kinesin motor clusters. **f)** The fundamental interactions of the primary components of the system are summarized. Filaments bundled together through depletion are crosslinked by Kinesin motor complexes.

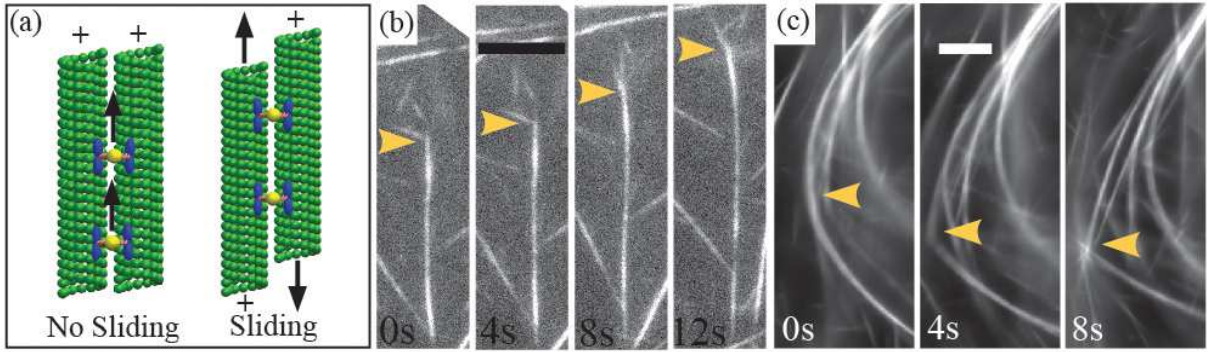

**Figure 2. Bundle extension and buckling. a**) When the MTs are polarity aligned kinesin clusters processively walk towards the plus end of the bundle without inducing MT sliding . Motor clusters will induce interfilament sliding MTs are polarity anti-aligned. **b**) Time sequence shows that a bundle of many MTs exhibits extensile behavior. Scale bar is 10µm. **c**) Bundles of MTs confined in a network become sterically hindered at their ends. A time sequence shows that this extensile force leads to bundle buckling. Scale bar is 10 µm.

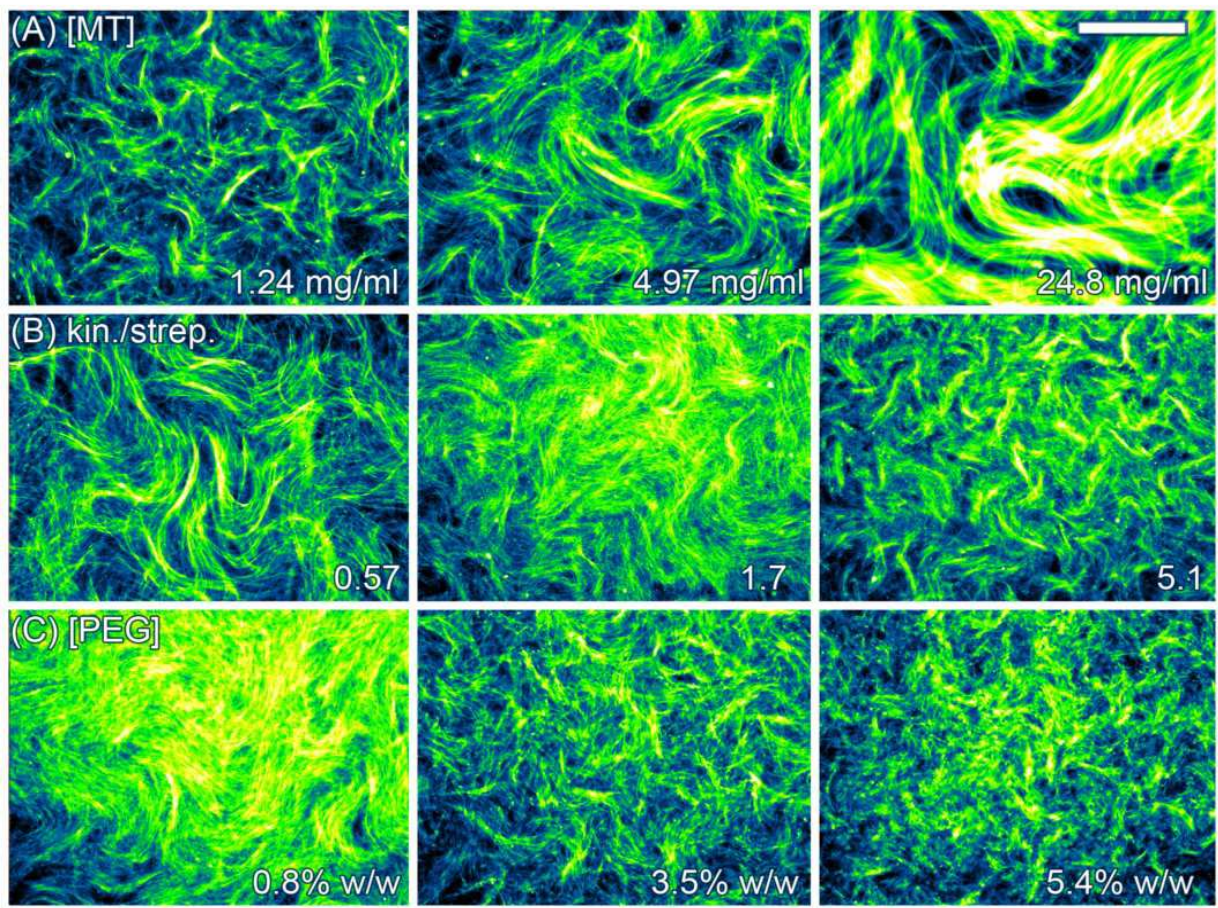

**Figure 3. Gel morphology from fluorescence microscopy. a)** Influence of increasing MT concentration on the structure of the active gel. **b)** Influence of increasing the ratio of kinesin to streptavidin on the structure of the active gel. **c)** Influence of increasing the PEG concentration on the structure of the active gel. All scale bars, 250 µm.

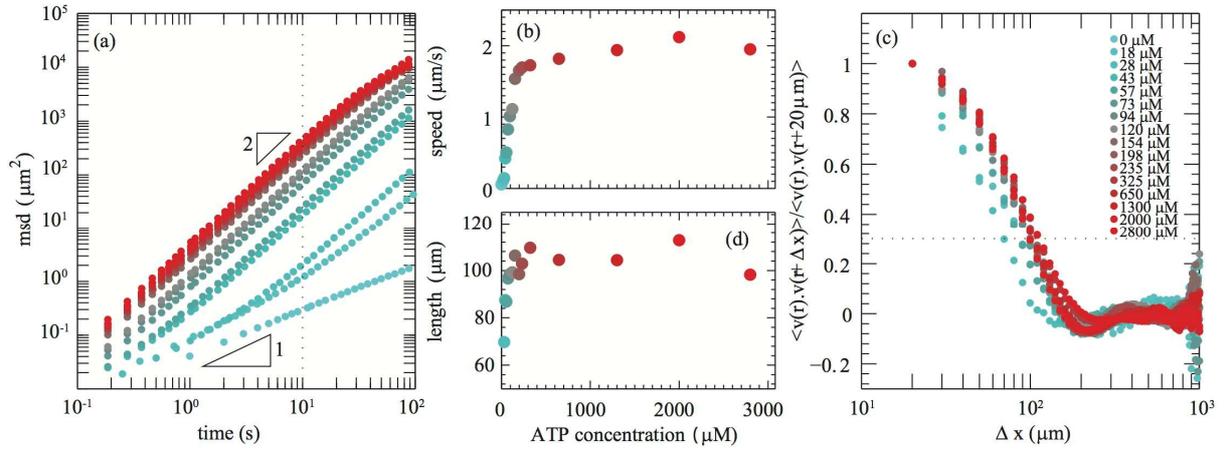

**Figure 4. Tracer particle dynamics for increasing ATP. a)** Dependence of tracer particle MSDs on ATP concentration. Reference guides for a slope of 1, indicating diffusive behavior, and a slope 2, indicating ballistic motion, are shown. The vertical dotted line indicates the time at which the characteristic speed of the tracer particles was measured. **b)** Characteristic speeds of tracer particles are plotted against the concentration of ATP. **c)** Normalized two-point spatial velocity-velocity correlations for tracer particles are plotted for increasing ATP concentrations. The horizontal dotted line indicates decay in correlation to 30% of the peak value and was used to measure λ the characteristic length scale of in the active gel. **d)** Dependence of λ on ATP concentration.

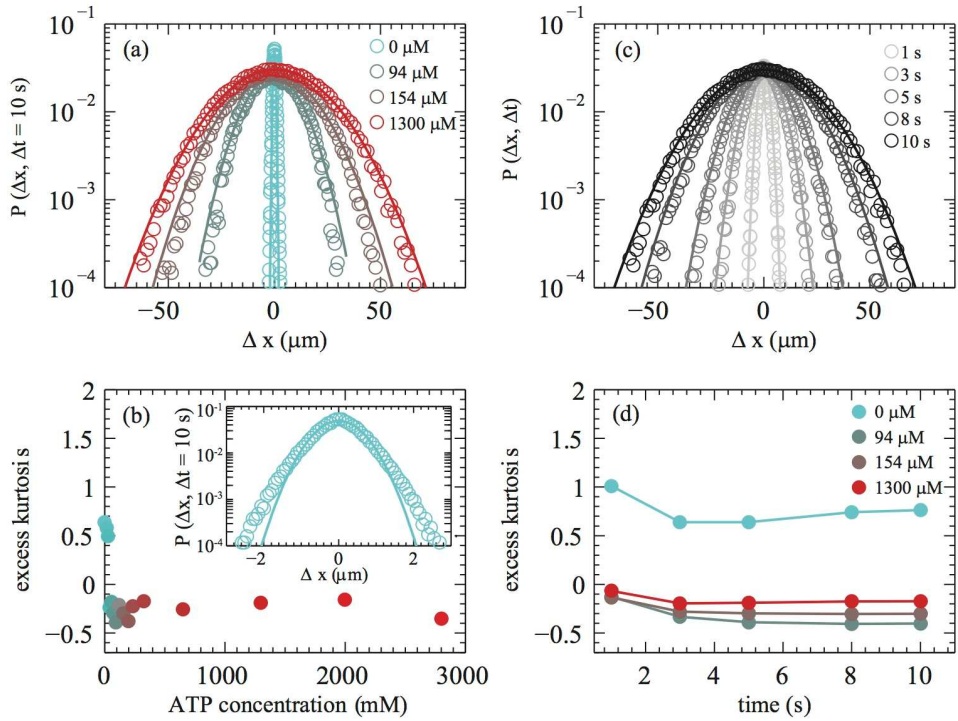

**Figure 5. Displacement PDF and excess kurtosis analysis for ATP.** a) Normalized PDFs of tracer bead displacements for varying ATP concentrations, over a time step of 10 seconds. b) Plot of excess kurtosis: $<u^4>/<u^2>^2 - 3$ over the range of ATP concentrations. A high kurtosis indicates a distribution that has broader tails than a Gaussian, as demonstrated in the inset graph: a blown up plot of the 0 ATP PDF at 10 seconds. c) Normalized PDFs of tracer bead displacements for ATP at 1.3 mM, in the saturation range, over varying time steps. d) Temporal evolution of excess kurtosis for a few values of ATP concentration, showing a positive value for the caged beads in the 0 ATP sample, and a negative kurtosis for active samples that tends to decrease with time step.

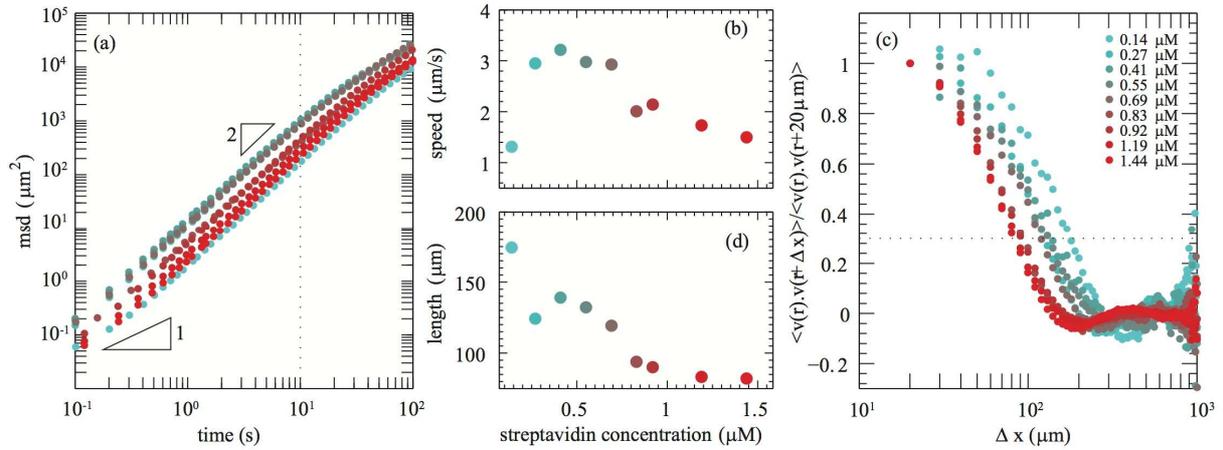

**Figure 6**. **Tracer particle dynamics for increasing kinesin motor cluster concentration. a)** MSDs of tracer particles plotted as a function of increasing motor cluster concentrations. **b)** Characteristic speeds of tracer particles are plotted against the motor cluster concentration. **c)** Normalized two-point spatial velocity-velocity correlations for tracer particles are plotted for increasing crosslinker concentrations. **d)** Dependence of λ on motor cluster concentration.

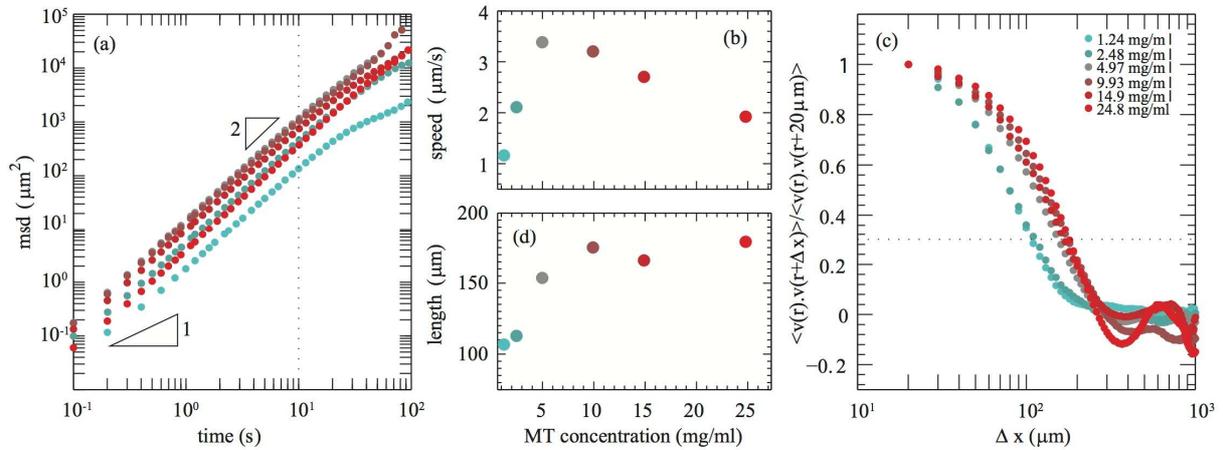

**Figure 7. Tracer particle dynamics for increasing MT concentration. a) Dependence of active gel** MSDs on MT concentrations. **b)** Characteristic speeds of tracer particles are plotted against the concentration of MTs. **c)** Normalized two-point spatial velocity-velocity correlations for tracer particles plotted for increasing MT concentrations. **d)** Dependence of λ on MT concentration.

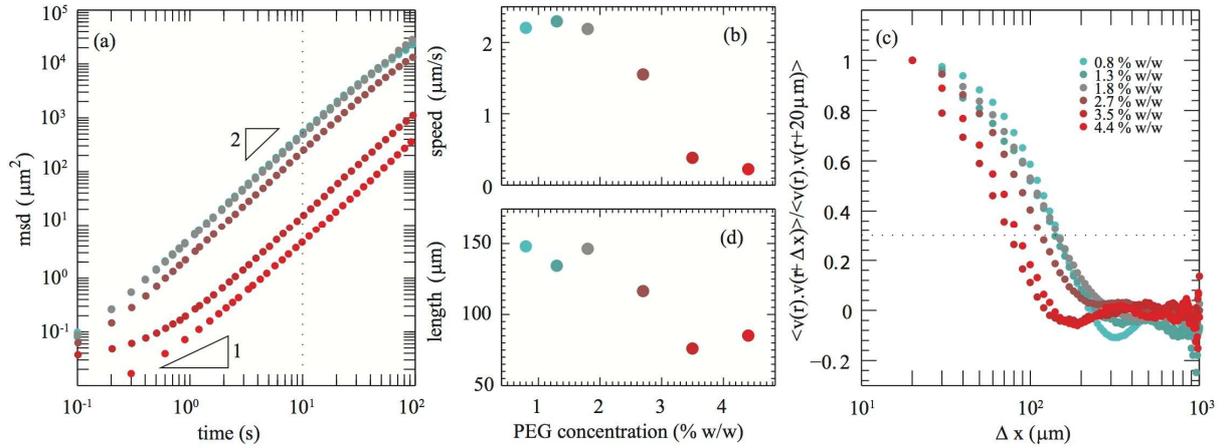

**Figure 8**. **Tracer particle dynamics for increasing PEG concentration. a)** MSDs displacements of tracer particles plotted against time for increasing PEG concentrations. **b)** Characteristic speeds of tracer particles are plotted against the concentration of PEG. **c)** Normalized two-point spatial velocity-velocity correlations for tracer particles plotted for increasing PEG concentrations. **d)** Dependence of λ on PEG concentration.

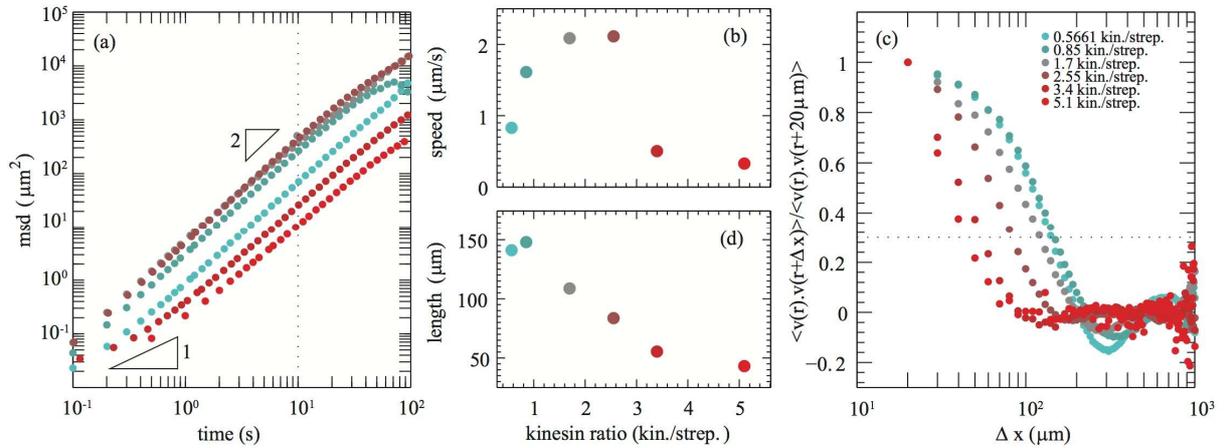

**Figure 9**. **Tracer particle dynamics for increasing the ratio of kinesin to streptavidin. a)** MSDs of tracer for active gels with increasing kinesin to streptavidin ratio. **b)** Characteristic speeds of tracer particles are plotted against the ratio of kinesin to streptavidin. **c)** Normalized two-point spatial velocity-velocity correlations for tracer particles plotted as fucntion of distance, for increasing ratio of kinesin to streptaavidin. **d)** Dependence of λ on kinesin to streptavidin ratio.